\documentclass[aps,prb,preprint,groupedaddress]{revtex4-1}
\usepackage[dvipdfmx]{graphicx}
\usepackage{mathrsfs}
\def\tc{$T_c$} \def\tl{$1/T_1$} \def\kb{$k_BT_c$} 

\begin{document}

\title{
Full-gap superconductivity in non-centrosymmetric Re$_6$Zr, Re$_{27}$Zr$_{5}$ and Re$_{24}$Zr$_{5}$}

\author{K. Matano$^{1}$, R. Yatagai$^1$, S. Maeda$^1$, and Guo-qing Zheng$^{1,2}$}
\affiliation{
$^1$Department of Physics, Okayama University, Okayama 700-8530, Japan\\
$^2$Institute of Physics, Chinese Academy of Sciences, and Beijing National Laboratory for Condensed Matter Physics, Beijing 100190, China
}
\date{\today}

\begin{abstract}
Non-centrosymmetric superconductor Re$_6$Zr has attracted much interest,
for its possible unconventional superconducting state with time reversal symmetry broken.
Here we report the $^{185/187}$Re nuclear quadrupole resonance (NQR) measurements on   Re$_6$Zr ($T_c$ = 6.72 K)
and the isostructural  compounds Re$_{27}$Zr$_{5}$ ($T_c$ = 6.53 K) and Re$_{24}$Zr$_{5}$ ($T_c$ = 5.00 K).
The nuclear spin-lattice relaxation rate $1/T_1$ shows a coherence peak below $T_c$  and
decreases exponentially at low temperatures in all three samples.
The superconducting gap $\Delta$ derived from the $1/T_1$ data  is $2\Delta =3.58 $ $k_BT_c$, 3.55  $k_BT_c$, and 3.51  $k_BT_c$
for Re$_6$Zr ,Re$_{27}$Zr$_{5}$, and Re$_{24}$Zr$_{5}$, respectively,
which is close to the value of 3.53 $k_BT_c$ expected for weak-coupling superconductivity.
These data suggest conventional $s$-wave superconductivity with a fully-opened gap in this series of compounds.
\end{abstract}


\maketitle
\section{introduction}
%
Superconductors with broken symmetries, such as broken time-reversal symmetry \cite{HillierQuintanillaCywinski2009} or broken spin-rotation symmetry \cite{MatanoKrienerSegawaEtAl2016}, have attracted great attention.
In particular, the role of crystal structure in the emergence of unconventional superconducting states   has been studied extensively  in recent years.

In superconductors  with an inversion center in the crystal structure,
either an even-parity spin-singlet or an odd-parity spin-triplet superconducting state is realized.
However, in non-centrosymmetric superconductors, a parity-mixed superconducting state is allowed
and an antisymmetric spin-orbit coupling (ASOC) interaction is induced\cite{Gorkov_Rashba_PhysRevLett.87.037004,Frigeri_PhysRevLett.92.097001,Frigeri_NJPhys.6.115}.
The parity-mixing extent is determined by the strength of the ASOC.

Indeed, some non-centrosymmetric superconductors show novel features.
For example, the isostructural Li$_2$Pd$_3$B and Li$_2$Pt$_3$B show contrasting behaviors.
Li$_2$Pd$_3$B exhibits conventional BCS-type properties\cite{Nishiyama_PhysRevB.71.220505},
while Li$_2$Pt$_3$B is a  spin-triplet dominant superconductor \cite{Nishiyama_PhysRevLett.98.047002} with nodes in the gap function
\cite{Nishiyama_PhysRevLett.98.047002,YuanAgterbergHayashi_PhysRevLett.97.017006}.
In this case, a different ASOC due to the difference of peculiar crystal structure distortion and atomic number
were responsible for the different superconducting states\cite{Harada_PhysRevB.86.220502}.
After the discovery of spin-triplet superconductivity in Li$_2$Pt$_3$B,
extensive studies have been performed to search for novel superconductivity in non-centrosymmetric superconductors
containing heavy elements such as Mg$_{10+x}$Ir$_{19-y}$B\cite{Klimczuk_Mg10Ir19B16_Phys.Rev.B.74.220502},
BiPd\cite{JoshiThamizhavelRamakrishnan_BiPd_Phys.Rev.B.84.064518}, and ScIrP\cite{OkamotoInoharaYamakawa_ScIrP_JPSJ.85.013704},
but parity-mixing is found to be weak
\cite{Tahara_PhysRevB.80.060503,matano_JPSJ.82.084711},
due to the crystal structure that does not lead to strong ASOC enhancement.

Recently, more novel properties were reported in some non-centrosymmetric superconductors.
For example, a small internal magnetic field was detected below \tc\ in non-centrosymmetric LaNiC$_2$,
which was interpreted as due to a breaking of time reversal symmetry in the superconducting state\cite{HillierQuintanillaCywinski2009},
although a relation between the breaking of inversion symmetry and time reversal symmetry is unclear.

Non-centrosymmetric Re$_6$Zr is a new candidate of time reversal symmetry-breaking superconductor.
Re$_6$Zr has an $\alpha$-Mn type cubic crystal structure with space group $I\bar{4}3m$\cite{Matthias_JPCS.19.130}
and a large upper critical field close to Pauli limit\cite{SinghHillierMazidian_Phys.Rev.Lett.112.107002}.
Figure 1 shows the $\alpha$-Mn type crystal structure, which has four independent crystallographic sites
Mn I : 2$a$ (Wyckoff), $\bar{4}3m$ (Hermann-Mauguin), Mn II :  8$c$, $3m$, Mn III : 24$g_1$, $m$, and Mn IV : 24$g_2$, $m$
\cite{HobbsHafnerSpisak_PhysRevB.68.014407}.
Among them, only the Mn I (2a, $\bar{4}3m$) site has an inversion center.
An internal magnetic field was detected in the superconducting state by muon spin relaxation or rotation ($\mu$SR) measurements\cite{SinghHillierMazidian_Phys.Rev.Lett.112.107002}.
The result was ascribed to  broken time reversal symmetry in the superconducting state.
It is known that  a chiral $p$-wave state or a chiral $d$-wave state can produce a tiny internal magnetic field
\cite{MackenzieMaeno_RevModPhys.75.657,Biswas_SrPtAs_PhysRevB.87.180503}.

In order to investigate the gap structure,
we performed the nuclear quadrupole resonance (NQR) measurements on 
 Re$_6$Zr, and the  iso-structural compounds Re$_{27}$Zr$_{5}$ and Re$_{24}$Zr$_{5}$.
These compounds have an $\alpha-$Mn type crystal structure but with different superconducting transition temperatures, with Re$_{24}$Zr$_{5}$ being stoichiometric.
The NQR measurement  performed at zero magnetic field  is  one of the most powerful methods for the study of the superconducting gap symmetry.
We find that all the three compounds show the superconducting  properties consistent with a conventional BCS gap symmetry.
\section{experimental}
The  polycrystalline samples of Re$_{6}$Zr, Re$_{27}$Zr$_{5}$, and Re$_{24}$Zr$_{5}$ in this study were synthesized by the arc-melting method.
The Re (99.99\%) and Zr (99.9\%) were arc-melted under an argon atmosphere.
The difference of mass before and after the melting was less then 1\% for all samples.
The melted ingots were crushed into powders for X-Ray diffraction (XRD) and NQR measurements.
The Cu K$_\alpha$ radiation is used for the XRD measurements.
The $T_c$ was determined by measuring the ac susceptibility using the {\it in situ} NQR coil.
A standard phase-coherent pulsed NMR spectrometer was used to collect data.
The nuclear spin-lattice relaxation rate was measured by using a single saturation pulse.
The spin echo was observed with a sequence of $\pi$/2 pulse (4 $\mu$s) - 30 $\mu$s - $\pi$ pulse (8 $\mu$s).
\begin{figure}[htbp]
\includegraphics[clip,width=80mm]{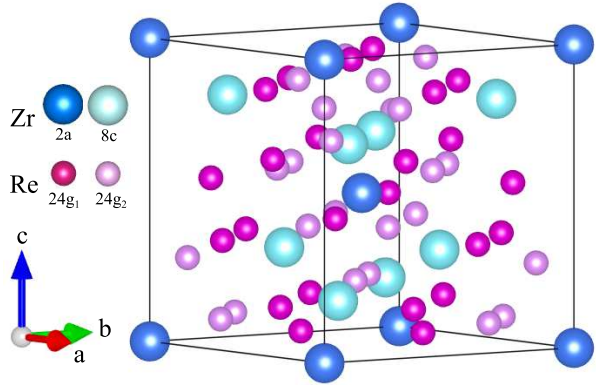}
\caption{\label{crystal}
(color online) Crystal structure of $\alpha$-Mn type Re-Zr system. It is a  cubic structure with space group $I\bar{4}3m$. }
\end{figure}

\begin{figure}[htbp]
\includegraphics[clip,width=74mm]{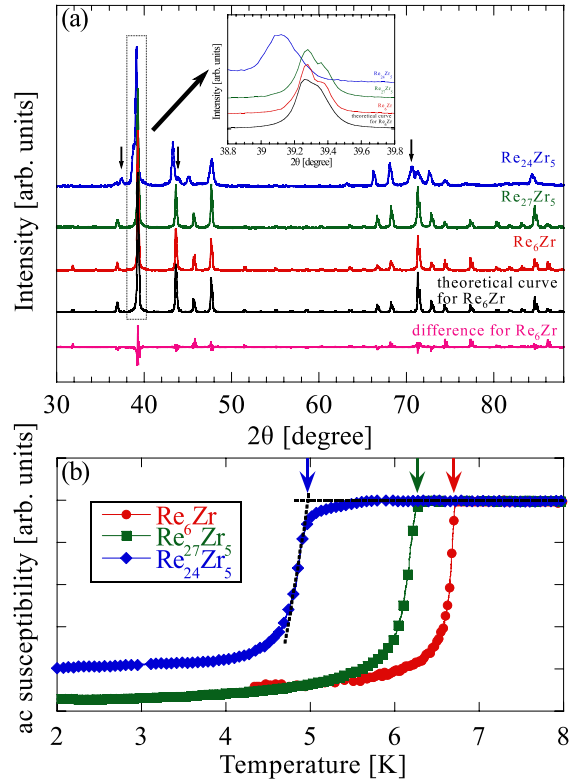}
\caption{\label{X-ray}(a) XRD patterns for Re$_{6}$Zr, Re$_{27}$Zr$_{5}$, and Re$_{24}$Zr$_{5}$.
The theoretical curve and the difference between the theoretical curve and the observed XRD are obtained by the Rietveld method.
Arrows indicate unidentified peaks.
For clarity, in the inset we show the enlarged part in the range of 38.8  - 39.8 degrees.
(b) Ac susceptibility measured using the $in$ $situ$ NQR coil at zero magnetic field. The arrows indicate \tc\ for each sample.}
\end{figure}

\section{results and discussions}
Figure \ref{X-ray} shows the XRD patterns and the ac susceptibility results for Re$_{6}$Zr, Re$_{27}$Zr$_{5}$ and Re$_{24}$Zr$_{5}$.
The lattice constant $a$ is 9.714 \AA\ for Re$_{6}$Zr, which is a little shorter than the reported value 9.698 \AA\cite{Matthias_JPCS.19.130}.
The \tc\ for Re$_{6}$Zr is 6.72 K, which is very close to 6.75 K reported in Ref. \cite{SinghHillierMazidian_Phys.Rev.Lett.112.107002}.
The obtained lattice constant and $T_c$ are shown in Table 1.    
For Re$_{6}$Zr and Re$_{27}$Zr$_{5}$, the XRD patterns can be fitted by Rietveld method.
For Re$_{24}$Zr$_{5}$, some unidentified peaks are observed.
The line width increases with increasing Zr composition, which suggests that impurities or crystal distortion increases with increasing Zr composition.
In the ac susceptibility for Re$_{24}$Zr$_{5}$,  a small shoulder can be seen around \tc,
which is likely attributable to the sample inhomogeneity.
Similar result has been reported in the Nb-Re systems\cite{ChenJiaoZhang_NdRe_PhysRevB.88.144510}.
%
In $\alpha$-Mn type systems, Re$_{24}$Zr$_{5}$ is stoichiometric, while
Re$_{6}$Zr and Re$_{27}$Zr$_{5}$ are non-stoichiometric with Re-rich compositions.
In the Re$_x$Zr$_{1-x}$ binary phase diagram \cite{Savitskii_Re_Zr},  a single phase  $\alpha$-Mn structure can be obtained only  in a narrow range with 0.82$\leq x \leq$0.87.
The Re-to-Zr ratio is  82.8 : 17.2  for the stoichiometric Re$_{24}$Zr$_{5}$, which is quite close to the limit to obtain a single phase.  On the other hand, the Re-to-Zr ratio is 84.4 : 15.6 for Re$_{27}$Zr$_{5}$ and 85.7 : 14.3 for Re$_{6}$Zr.
As seen in Fig. 2,   the stoichiometric Re$_{24}$Zr$_{5}$ compound showed a small amount of additional peaks in the XRD chart, which is
 likely due to the fact that the Re$_{24}$Zr$_{5}$ is close to the phase boundary. 


\begin{table*}[t]
\caption{\label{}Crystal structure, the NQR parameters for $^{187}$Re, and the superconductivity parameters for Re$_6$Zr, Re$_{27}$Zr$_{5}$, and Re$_{24}$Zr$_{5}$.
  }
\begin{ruledtabular}
\begin{tabular}{rcccccc}

 &\multicolumn{2}{c}{Re$_6$Zr (Re$_{30}$Zr$_{5}$)}& \multicolumn{2}{c}{Re$_{27}$Zr$_{5}$}&\multicolumn{2}{c}{ Re$_{24}$Zr$_{5}$} \\
 lattice constant [\AA] & \multicolumn{2}{c}{9.714} &\multicolumn{2}{c}{9.726}&\multicolumn{2}{c}{9.762}\\
  $2\Delta$ [$k_BT_c$]&\multicolumn{2}{c}{3.58} &\multicolumn{2}{c}{3.55}&\multicolumn{2}{c}{3.51}\\
 $T_c$ [$K$]& \multicolumn{2}{c}{6.72} & \multicolumn{2}{c}{6.53} &\multicolumn{2}{c}{5.00} \\\hline
site& A & B & A & B & A & B\\\hline
$\nu_Q$ [MHz]& 42& 84& 42&83&42&75\\
 $\eta$ &0.6&0.6 &0.6&0.6 &0.6&0.7\\
 $FWHM$ ($\pm1/2\leftrightarrow\pm3/2$)  [MHz]&13&25 &14&20 &16&23\\
 $FWHM$ ($\pm3/2\leftrightarrow\pm5/2$)  [MHz]&19&36 &24&34 &28&37\\

\end{tabular}
\end{ruledtabular}
\end{table*}

\begin{figure}[htbp]
\includegraphics[clip,width=74mm]{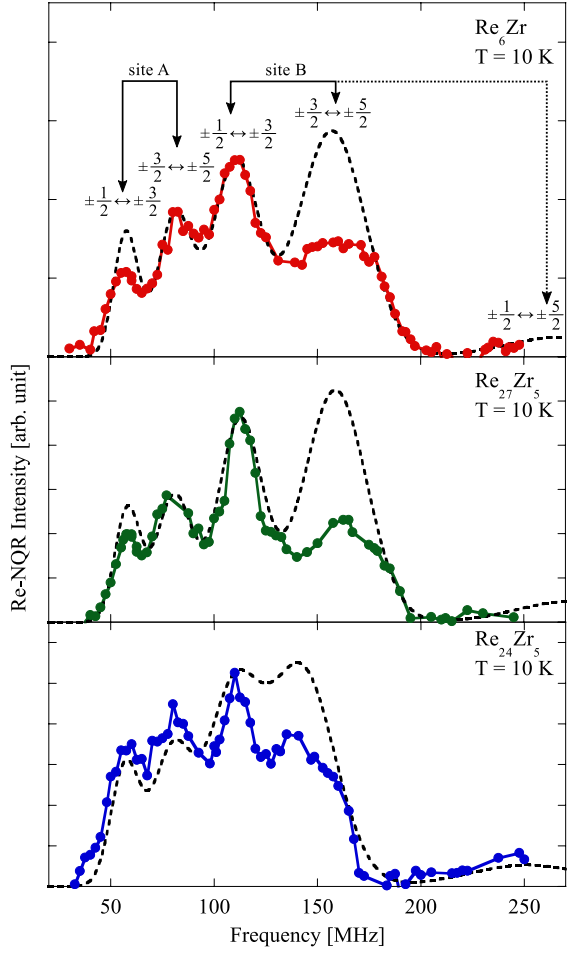}
\caption{\label{f1}$^{185/187}$Re-NQR spectrum of Re$_{6}$Zr, Re$_{27}$Zr$_{5}$ and Re$_{24}$Zr$_{5}$ measured at $T$ = 10 K.
Dotted curves are the theoretical simulations (see text).}
\end{figure}

Figure \ref{f1} shows the $^{185/187}$Re-NQR  spectra at $T$ = 10 K for the three samples.
Four peaks were observed for all samples.
The Re nuclei have a spin $I = \frac{5}{2}$, which will result in two NMR transitions.
In the Re-Zr systems with $\alpha-$Mn type structure,
the unit cell has 58 atoms that are distributed into two Zr sites ($2a$, $8c$) and two Re sites (24$g_1$, 24$g_2$)\cite{JoubertProg.Mate.Sci.54.945}.
Furthermore, Re has two isotopes $^{185}$Re (natural abundance 37.5\%) and $^{187}$Re (62.5\%).
As a result, eight peaks for this compound are expected in principle.
However, the difference in the nuclear quadrupole moment $Q$ of $^{185}$Re ($2.7\times 10^{-24} $ cm$^2$) and  $^{187}$Re ($2.6\times 10^{-24} $ cm$^2$) is only 4\%,
which lead to the inability of distinguishing $^{185}$Re from $^{187}$Re in the broad spectra and thus only four peaks were observed.
As can be seen in Fig. 3, only the uppermost peak varies upon changing the Re-Zr composition ratio.
By a theoretical simulation (see below), we  assigned the lower two peaks to site A   and the upper two peaks to site B. At the moment, it is unclear site A (B) corresponds to which Re site in the crystal structure.
%
The Hamiltonian for the quadrupole interaction  is,
\begin{equation}
\mathscr{H} = \frac{\nu_Q}{6} \left\{(3I^2-\vec{I}^2)+\frac{\eta}{2}(I_{+}^2+I^2_{-})\right\}.
\end{equation}
Here $\nu_Q$ and $\eta$ are defined as
\begin{eqnarray}
\nu_Q \equiv \nu_z =\frac{3}{2I(2I-1)h}e^2Q\frac{\partial^2 V}{\partial z^2}\\
\eta = \frac{|\nu_x-\nu_y|}{\nu_z},
\end{eqnarray}
with $\frac{\partial^2 V}{\partial \alpha^2} (\alpha = x, y, z) $
being the electric field gradient at the position of the nucleus.
In the simulations, a Gaussian function $\exp \left\{-(f/2\delta)^2 \right\}$ is convoluted,
where $f$ is the frequency and $\delta$ is related to the full width at half maximum ($FWHM$) of the transition line as $FWHM = 2\delta\sqrt{2\ln(2)}$.
The $\nu_Q$ and  $\eta$ were treated as parameters.
The intensity ratio of each peak depends on $\eta$. With the
 parameters  listed in Table 1, we are able to reproduce the experimental results as seen in Fig. 3. We note that  $\nu_Q$ or the FWHM is proportional to $Q$,
so the value for $^{185}$Re ($Q= 2.7\times 10^{-24} $ cm$^2$) is 1.04 times the value for $^{187}$Re  ($Q = 2.6\times 10^{-24} $ cm$^2$).
The intensity for the uppermost peak does not agree with the simulation,
probably due to a worse quality factor of the coil in this frequency range.
%
%
In the case of large $\eta$, signal from the forbidden transitions ($\pm1/2\leftrightarrow\pm5/2$) is expected.
Indeed, we detected such signals  in the frequency range above 200 MHz as seen in Fig. 3.
\begin{figure}[htbp]
\includegraphics[clip,width=70mm]{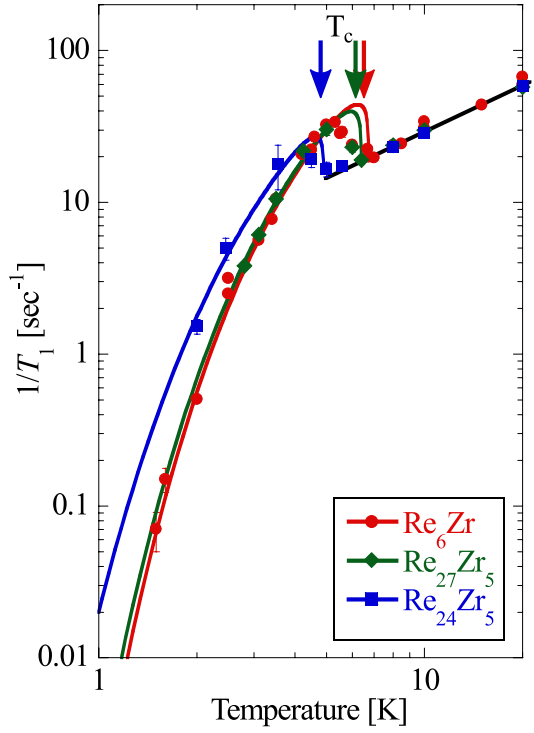}
\caption{\label{T1}
(color online) Temperature dependence of the spin-lattice relaxation rate $1/T_1$ measured by NQR.
The straight line above \tc\ represents the $T_1T$ = const relation.
The solid curve below \tc\ is a calculation assuming the $s$-wave gap function (see text).
}
\end{figure}


Figure \ref{T1} shows the temperature dependence of \tl\ of $^{185/187}$Re-NQR,
which was measured at the 1$\nu_Q$ ($\pm1/2\leftrightarrow\pm3/2$) transition of the site B.
The nuclear magnetization decay curve for each compound is well fitted to the theoretical formula for different $\eta$\cite{Chepin_eta_not_0},
\begin{eqnarray}
{\scriptstyle {\rm Re_{6}Zr}\ :\ \frac{M_0-M(t)}{M_0}} &=
{\scriptstyle 0.163\exp\left(-\frac{3.00t}{T_1}\right)+
 0.837\exp\left({-\frac{8.52t}{T_1}}\right)}\\
{\scriptstyle{\rm Re_{27}Zr_{5}} \ :\ \frac{M_0-M(t)}{M_0}}&=
{\scriptstyle  0.170\exp\left({-\frac{3.00t}{T_1}}\right)+
0.830\exp\left({-\frac{8.42t}{T_1}}\right)}\\
{\scriptstyle{\rm Re_{24}Zr_{5}} \ :\ \frac{M_0-M(t)}{M_0}}&=
{\scriptstyle 0.187\exp\left({-\frac{3.00t}{T_1}}\right)+
0.813\exp\left({-\frac{8.23t}{T_1}}\right)}
\end{eqnarray}
where $M_0$ is the nuclear magnetization in the thermal equilibrium and $M(t)$ is the nuclear
magnetization at a time $t$ after the saturating pulse.
We have confirmed that $T_1$ measured at the $\pm3/2\leftrightarrow\pm5/2$ transition gives the same value.
The recovery curve can be fitted with a single $T_1$  component below and above \tc,
which indicates that macroscopic phase separation in the sample is small.
As seen in the figure, \tl\ varies in proportion to the temperature ($T$) above \tc\ for all samples, as expected for conventional metals, indicating no electron-electron interaction.
Below \tc, \tl\ shows a coherence peak (Hebel-Slichter peak) expected for an $s$-wave superconducting state.
The $1/T_{1S}$ in the superconducting state is expressed as
\begin{eqnarray}
\frac{T_{1N}}{T_{1S}}=\frac{2}{k_BT}
\int\left(1+\frac{\Delta^2}{EE'}\right)N_S(E)N_S(E')  \nonumber \\
\times f(E)\left[1-f(E')\right]\delta(E-E')dEdE',
\end{eqnarray}
where $1/T_{1N}$ is the relaxation rate in the normal state, $N_S(E)$ is the superconducting density of
states (DOS), $f(E)$ is the Fermi distribution function and $C =1+ \frac{\Delta^2}{EE'}$ is the coherence factor.
To perform the calculation of eq. (6), we follow Hebel to convolute $N_S(E)$ with a broadening function $B(E)$\cite{Hebel},
which is approximated by a rectangular function centered at $E$ with a height of 1/2$\delta$.
The solid curve below \tc\ shown in Fig. 4  is a calculation with
$2\Delta = 3.58$\kb, $b \equiv \delta/\Delta(0)=0.030$ for Re$_{6}$Zr,
$2\Delta = 3.55$\kb, $b =0.058$ for Re$_{27}$Zr$_{5}$, and
$2\Delta = 3.51$\kb, $b =0.107$ for Re$_{24}$Zr$_{5}$.
The curve fits the experimental data reasonably well. The parameter $2\Delta$ is close to the BCS value of 3.53\kb.
This result indicates an isotropic superconducting gap in these compounds.
A similar conclusion was drown by a resent London penetration depth measurement for Re$_{6}$Zr\cite{KhanKarkiPrestigiacomo_Re6_Zr_arXiv1603.07297}.

Our result is inconsistent with a time reversal symmetry
broken superconducting state such as $d + id$ or $p + ip$ where the coherence peak will be absent.
We note that inconsistent results from different probes have been reported in LaNiC$_2$, PrPt$_4$Ge$_{12}$, and
the locally non-centrosymmetric superconductor SrPtAs that has an inversion center in whole unit cell but not within a single layer.
In these samples, time reversal symmetry breaking was suggested by $\mu$SR
\cite{HillierQuintanillaCywinski2009,Maisuradze_PrPt4Ge12_uSR_PhysRevB.82.024524,Biswas_SrPtAs_PhysRevB.87.180503},
but an $s$-wave superconductivity was confirmed by NQR/NMR measurements
\cite{Iwamoto_LaNiC2_Phys.Lett.A.250.439,Kanetake_PrPt4Ge12_JPSJ.79.063702,Matano_SrPtAs_PhysRevB.89.140504}.
In these materials, breaking of time reversal symmetry has not been observed except for $\mu$SR.
\section{summary}
In summary, we have performed the $^{185/187}$Re-NQR measurements on the
non-centrosymmetric superconductors Re$_{6}$Zr, Re$_{27}$Zr$_{5}$, and Re$_{24}$Zr$_{5}$.
The $T$-linear behavior of the nuclear spin-lattice relaxation rate $1/T_1$ above $T_c$ indicates the absence of spin correlations.
The $1/T_1$ shows a Hebel-Slichter peak just below $T_c$ and decreases exponentially at low temperatures for all samples,
which suggests that the $\alpha-$Mn type Re-Zr system is in an $s$-wave fully-gapped superconducting state,
which is inconsistent with a time reversal symmetry breaking state such as $d+id$ or $p+ip.$
\begin{acknowledgments}
This work was supported by Research Grants,
 No. 15H05852 (Innovative Areas ``Topological Materials Science'') from MEXT
and No. 16H04016 from JSPS.
\end{acknowledgments}

\end{document}